\documentclass[12pt,twoside,fleqn]{article}
\usepackage[dvips]{epsfig}
\usepackage{latexsym}

\def\lsim{\raise0.3ex\hbox{$<$\kern-0.75em\raise-1.1ex\hbox{$\sim$}}}
\def\gsim{\raise0.3ex\hbox{$>$\kern-0.75em\raise-1.1ex\hbox{$\sim$}}}
\newcommand{\dd}{\dagger}

\newcommand{\half}{\frac{1}{2}}

\newcommand{\tr}{\mbox{tr}}

\newcommand{\ld}{\lambda}

\newcommand{\are}{adjoint representation}

\newcommand{\bd}{\beta_{\rm d}}
\newcommand{\bc}{\beta_{\rm c}}
\newcommand{\td}{T_{\rm d}}
\newcommand{\tc}{T_{\rm c}}
\newcommand{\ie}{{\sl i.e.~\/}}

\begin{document}
\thispagestyle{empty}
%\title{
 \mbox{} \hfill BI-TP 98/40\\
%\mbox{} \hfill {hep-lat/98xxxxx} \\
 \mbox{} \hfill December 1998\\
\begin{center}
{{\large \bf
Deconfinement and Chiral Symmetry Restoration\\
in an SU(3) Gauge Theory with Adjoint Fermions
\bigskip
} \\
\vspace*{1.0cm}}
%\author{
Frithjof Karsch and Martin L\"utgemeier

\vspace*{1.0cm}
{\normalsize
$\mbox{}$ {Fakult\"at f\"ur Physik, Universit\"at Bielefeld,
D-33615 Bielefeld, Germany}
}
\end{center}
\vspace*{1.0cm}
\centerline{\large ABSTRACT}

%\maketitle

%\begin{abstract}
We analyze the finite temperature phase diagram of $QCD$ with fermions in 
the adjoint representation. The simulations performed with four dynamical
Majorana fermions show 
that the deconfinement and chiral phase transitions occur at two distinct 
temperatures. While the deconfinement 
transition is first order at $\td$ we find evidence for a continuous chiral
transition at a higher temperature $\tc \simeq 8~\td$. 
We  observe a rapid change of bulk thermodynamic observables 
at $\td$ which reflects
the increase in the number of degrees of freedom. However, these show little
variation at $\tc$, where the fermion condensate vanishes. We also analyze 
the potential between static fundamental and adjoint charges in all three phases
and extract the corresponding screening masses above $\td$. 

%\end{abstract}selineskip 20pt

\noindent

\vskip 20pt
\vfill
\eject
\baselineskip 15pt

\section{Introduction}
The interplay between confinement and chiral symmetry restoration is one
of the most puzzling problems in finite temperature $QCD$.
Both phenomena seem to refer to different non perturbative mechanisms, where
a priori unrelated length scales are involved.
But speculations, that $QCD$ would therefore undergo two distinct phase
transitions \cite{Shuryak} -  deconfinement of quarks and gluons at $\td$ 
followed by the restoration of chiral symmetry at some higher 
temperature $\tc$ - could not be confirmed by lattice calculations.
The two effects seem to be tightly coupled in $QCD$.
At a unique critical temperature the chiral condensate vanishes and the
asymptotic value of the heavy-quark potential drops rapidly indicating
deconfinement. 

Thermodynamic quantities calculated in the vicinity of the $QCD$
phase transition temperature exhibit the deconfining as well as
chiral symmetry restoring aspects of this transition. Observables
like the energy or entropy density show a sudden rise at $\tc$ which
reflects the liberation of many new partonic degrees of freedom due to
deconfinement. In this respect the transition is very similar to that
in the pure gauge sector where quarks are absent and chiral symmetry thus
does not play a role. In $QCD$ with $n_f$ light quark flavours, however, also
the chiral condensate drops rapidly and vanishes at $\tc$
for vanishing quark masses. In fact, it generally is believed that details 
of the $QCD$ phase transition, e.g. the order of the transition, is controlled
by chiral symmetry. At least qualitatively this has been confirmed
by numerical calculations. While the transition is first order
for $QCD$ with three or more light flavours it does seem to be continuous
for two flavours as expected from general universality arguments 
\cite{Pisarski},
although at present the analysis of critical exponents for two 
flavour ($n_f=2$) $QCD$ 
is not entirely in agreement with expectations based on the  universality class
of 3-dimensional $O(4)$ spin models \cite{KaL,Tsukuba}. 
Also characteristic chiral singularities, which are expected to show up in 
the quark mass dependence of the chiral susceptibility
for $T\lsim T_c$ \cite{Zia} have so far not been observed.

In how far are the chiral properties of the transition dependent on the 
confinement-deconfinement transition and in how far is chiral symmetry
restoration reflected in the behaviour of bulk thermodynamic observables? 
In order to get further insight into these questions 
one may analyze $QCD$ related models in which both transitions fall apart. 
The  $SU(N)$ gauge theories with fermions in the adjoint representation ($aQCD$)
offer this possibility. Early investigations of $SU(2)$ with 
two\footnote{In the following
we will adopt the $QCD$ convention when counting the number of fermion flavours,
i.e. we count the number of Dirac fermion species in the adjoint 
representation rather than Majorana fermions.} 
adjoint fermions \cite{kogut85,kogut87} 
indeed suggest that the two phase transitions are 
well separated with $\td<\tc$. This may not be too surprising as $aQCD$ 
possesses two distinct global symmetries. 
Unlike fermions in the fundamental representation the adjoint 
fermions do not break the global $Z(N)$ center symmetry of the $SU(N)$ gauge group.
The model thus has 
the $Z(N)$ center symmetry and a $SU(2n_f)$ chiral symmetry \cite{peskin,smilga}.
At low temperatures the latter is broken down to $SO(2n_f)$ while the former is 
expected to become broken at high temperature.
Related to these symmetries and their breaking or restoration are 
two order parameters, the Polyakov loop expectation value and the expectation
value of the fermion 
condensate, which can characterize the different phases of the 
theory\footnote{We note that the model we are discussing here is closely related
to supersymmetric gauge theories. For a recent attempt to simulate
supersymmetric Yang Mills theories on the lattice see \cite{Montvay}.
The possibility of the breaking of
the $Z(N)$ center symmetry in finite temperature supersymmetric models
has recently also been discussed in Ref. \cite{witten}.}.

In this paper we concentrate on an analysis of the thermodynamics of $aQCD$.
We determine the critical
temperatures, analyze the order of the transitions and give an estimate
of the latent heat at the deconfinement transition. Particular emphasis
is put on the investigation of the thermodynamics in the  
intermediate phase, {\it i.e.} for $\td < T < \tc$,
where we analyze
the temperature and fermion mass dependence of the fermion condensate,
its susceptibility as well as 
potentials for static fermions in the fundamental and adjoint representations.
Before discussing the thermodynamics of $aQCD$ in section 3 we will
introduce the model and give 
a description of numerical aspects of our calculations in section 2.
Section 4 contains our conclusions.

\section{SU(3) gauge theory with fermions in the adjoint representation: 
${\bf aQCD}$}

In this section we want to introduce the basic properties of the lattice
formulation of the $SU(3)$ gauge theory with fermions in the {\are}.
We will use for our investigations the standard Wilson one-plaquette
gauge action. In the fermion sector we use the staggered fermion
formulation. The only modification compared to $QCD$ calculations is 
that the fermions are in the adjoint (8-dimensional) representation
of the $SU(3)$ group. As they carry now eight colour degrees of
freedom we also need an 8-dimensional representation for the gluon
fields in the discretization of the fermionic action, 

\begin{equation}
\label{eq:actiona}
  S = \beta S_G
  + \sum_{x,y} \bar{\psi}_x  M(U^{(8)})_{x,y} \psi_y \quad,
\end{equation}
where $M(U^{(8)})$ denotes the standard staggered fermion matrix with the 
3-dimensional gauge field matrices $U^{(3)}$ replaced  by the 8-dimensional
representation  $U^{(8)}$, $\beta=6/g^2$ is related to the inverse coupling
of the gauge fields  and $S_G$ denotes the gluonic part of the action,
\begin{equation}
\label{eq:gluonaction}
S_G= \sum_{\Box} \left\{ 1 - \frac{1}{3} {\rm Re Tr} U_{\Box}^{(3)} \right\}
\quad .
\end{equation}
We note that the  8-dimensional representation may be represented in terms of 
the 3-dimensional one as
\begin{equation}
  \label{eq:adjoint}
  U^{(8)}_{ab} = \half\tr[\ld_aU^{(3)}\ld_b{U^{(3)}}^\dd]\quad ,
\end{equation}
with $\ld_a$ being the Gell-Mann matrices and $\tr$ the 3-trace. This also 
shows explicitly that the action for adjoint fermions does not break the $Z(3)$
center symmetry. In the fermionic part of the action the matrices $U^{(3)}$  
appear only in  complex conjugate pairs
$U^{(3)}$, $U^{(3)^{\dd}}$. The two center elements appearing in a $Z(3)$
transformation therefore cancel each other.
Like in the $SU(3)$ gauge theory the Polyakov loop, 
\begin{equation}
  \label{eq:fpoly}
L_3 \equiv \lim_{ N_\sigma \rightarrow \infty}\langle  {1\over N_\sigma^3 }  
\bigl|\sum_{\vec{x}} L_3(\vec{x})\bigr|\rangle  = 
\lim_{ N_\sigma\rightarrow \infty}\langle 
{1\over 3 N_\sigma^3 } \bigl|\sum_{\vec{x}} {\rm Tr} \prod_{x_0=1}^{N_\tau} 
U^{(3)} (x_0,\vec{x}) \bigr|\rangle\quad,
\end{equation}
thus still is an order 
parameter of the theory for the spontaneous breaking of the center symmetry.
As its value is related to the asymptotic behaviour of the Polyakov loop
correlation function, 
\begin{equation}
  \label{eq:fpotential}
\exp(-V_3 (\vec{x},T)/T) \equiv \langle L_3(\vec{0}) L_3^{\dagger} (\vec{x})\rangle
\quad_{\longrightarrow \atop {|\vec{x}|\rightarrow \infty}} \quad L_3^2 \quad , 
\end{equation}
which defines the potential between static fundamental charges, it is obvious
that the thermal medium generated by $aQCD$ will be confining for fundamental
charges at low temperatures as long as the $Z(3)$ center symmetry is not 
spontaneously broken. In contrast to this the expectation value of the 
adjoint Polyakov loop\footnote{To evaluate adjoint traces we may use the
relation ${\rm Tr}U^{(8)} = |{\rm Tr}U^{(3)}|^2-1$, which can easily be derived 
from equation \ref{eq:adjoint}.},
\begin{equation}
  \label{eq:apoly}
L_8 \equiv  
\lim_{ N_\sigma\rightarrow \infty}\langle 
{1\over N_\sigma^3 }  
\sum_{\vec{x}} L_8(\vec{x})\rangle = 
\lim_{ N_\sigma\rightarrow \infty}\langle 
{1\over 8 N_\sigma^3 } \sum_{\vec{x}} {\rm Tr} \prod_{x_0=1}^{N_\tau} 
U^{(8)} (x_0,\vec{x}) \rangle \quad, 
\end{equation}
is non-zero for all values of $\beta$. Adjoint charges are thus screened
and the corresponding potential reaches a constant value at large 
distances\footnote{Note that this screening is not only related to the fact 
that adjoint charges appear now
as dynamical degrees of freedom in $aQCD$. Even in the quenched case, \ie the 
pure $SU(3)$ gauge theory, adjoint charges are screened at low temperatures. 
String breaking has been analyzed in this limit \cite{stringb}.},
\begin{equation}
  \label{eq:apotential}
\exp(-V_8 (\vec{x},T)/T) \equiv \langle L_8(\vec{0}) L_8 (\vec{x})\rangle
\quad_{\longrightarrow \atop {|\vec{x}|\rightarrow \infty}} \quad L_8^2 \quad .  
\end{equation}

We note that the matrices $U^{(8)}$ are real and so is the entire 
fermion matrix $M$ and its determinant. As a consequence of this pseudo 
fermion fields $\Phi$, which are generally used to represent the fermion 
determinant as a Gaussian integral over bosonic fields, can be chosen
real. This reduces the number of fermion doublers by a factor of 2, {\it i.e.}
we can set up an exact fermion algorithm for two adjoint fermion flavours
(see footnote a).

The pseudo fermion action, used in our simulations thus reads
\begin{equation}
  \label{eq:action}
  S = \beta \sum_{\Box} \left\{ 1 - \frac{1}{3} {\rm Re Tr} U_{\Box}^{(3)} \right\}
  + \Phi^t \left\{ M(U^{(8)})^t M(U^{(8)}) \right\}^{-1} \Phi \quad.
\end{equation}

For our calculations we use the hybrid Monte Carlo algorithm.
We note that the entire molecular dynamics evolution of the system can be 
expressed in terms of the 3-dimensional matrices $U^{(3)}$ and the 
fermion fields. An explicit representation of the matrices $U^{(8)}$ 
is only needed in the conjugate gradient routine used for the inversion of the
fermion matrix. 
Technically the largest problem this prescription causes is the evaluation
of the time derivative of $M^{t}M$, which is needed in the molecular 
dynamics steps. As the fermion matrix now is quadratic in $U^{(3)}$
time derivatives generate also twice as many terms in the Hamiltonian 
evolution equations as in ordinary $QCD$ simulations.

Our simulations were performed on  lattices of size $N_\sigma ^3 \times N_\tau$
with $N_\tau=4$ and $N_\sigma =8$ and 16.
On the smaller lattice  we performed simulations for 15 $\beta$-values in 
the interval $[5.2;6.5]$ each for the bare masses $m= 0.02$, 0.04, 0.08, 0.10.
In some cases additional runs were performed at a smaller fermion mass,
$m=0.01$, as well as at intermediate masses, $m=0.03$ and $m=0.05$.
On the larger lattice calculations were performed with the fermion
mass $m=0.02$ for 8 $\beta$-values between 5.25 and 6.5. 
At some selected values of the gauge coupling, $\beta=5.25, 5.4, 6.2, 6.5$,
which were chosen in the three different phases of thermal $aQCD$, we also 
calculated the fundamental and adjoint Polyakov loop correlation functions
and extracted the corresponding potentials from them.

We used the hybrid-$\Phi$ algorithm \cite{got} with the standard gluon and staggered
fermion action only modified for the {\are}, and the conjugate gradient method
for the inversion of the fermion matrix.
The residuum was $10^{-8}$ inside the molecular dynamic steps and $10^{-13}$
elsewhere. The length of a trajectory has been fixed to $\Delta t = 0.25$. 
The number of subdivisions of a trajectory has been chosen such that we 
reached an acceptance rate of $60-80\%$ which could drop to $50\%$ in the 
vicinity of $\bd$. It turned out that the number of subdivisions needed mainly
depends on the fermion mass and shows little dependence on $\beta$. 
More details on our simulation parameters are given in Table \ref{tab:nmd}.

\begin{table}[htbp]
  \begin{center}
    \leavevmode
    \begin{tabular}[t]{|l|c|c|c|}
      \hline
      mass \rule[-1ex]{0em}{3.4ex} & \#mol.dyn. steps & \#traj. for therm. & \#traj. for measurement \\
      \hline
      \multicolumn{4}{l}{$8^3\times 4$ lattice \rule{0em}{4ex}} \\
      \hline
      0.01 & 64    & 500        & 1000 - 3000 \\
      0.02 & 40    & 500        & 1500 - 6500 \\
      0.03 & 32    & 500        & 1500        \\
      0.04 & 28    & 500        & 2800 - 7500 \\
      0.05 & 25    & 500        & 2500 - 6500 \\
      0.08 & 21    & 500 - 1000 & 1500 - 7000 \\
      0.10 & 20    & 500 - 1000 & 400 - 14000 \\
      \hline
      \multicolumn{4}{l}{$16^3\times 4$ lattice \rule{0em}{4ex}} \\
      \hline
      0.02 & 50-75 & 100 - 1000 &  500 - 2300 \\
      \hline
    \end{tabular}
  \end{center}
  \caption{Parameters used in the hybrid Monte Carlo calculations on different
lattice sizes and for different values of the fermion mass.}
  \label{tab:nmd}
\end{table}

Plaquette and Polyakov loop expectation values have been calculated 
on every configuration, the
fermion condensate only on every 5th for the small lattice and every second
configuration for the larger lattice. On these configurations we also
calculated the disconnected part of the fermionic susceptibility \cite{KaL}, 
\begin{equation}
  \label{eq:chi}
\chi_m = {1 \over N_\sigma^3 N_\tau}\biggl(
\langle ({\rm Tr} M^{-1})^2\rangle - \langle {\rm Tr} M^{-1}\rangle^2
\biggr)
\end{equation}
for which we  used a noisy estimator \cite{Bitar} with 25 random source vectors. 

Correlation functions of Polyakov loops for fundamental and adjoint
static fermions have been calculated in order to extract the potential
for both types of fermions. The correlators were measured every second
iteration at distances $R=\sqrt{x_1^2+x_2^2+x_3^2}$, $x_i=0,1,2,3,4$ and 
also at on-axis distances 5, 6, 7 and 8. After averaging over correlations 
at equal distance we obtained the
potentials at 33 points for $1.0\le R \le  8.0$.

\section{Thermodynamics of $aQCD$}

As discussed in the previous section the Polyakov loop, $L_3$, is an order
parameter for the spontaneous breaking of the $Z(3)$ center symmetry of
$aQCD$ and easily is related to the confining/deconfining properties of the
static potential for fundamental charges (Eq.~\ref{eq:fpotential}). Less
obvious is that this also reflects the deconfinement of the dynamical 
degrees of freedom of $aQCD$, \ie the adjoint fermions. In order to analyze 
this question we have calculated various observables reflecting thermodynamic
properties of the hot medium which we will discuss in the following.
We will start with a discussion of signals for the deconfinement and chiral
transitions and then discuss thermodynamic observables related to the 
equation of state of $aQCD$.

\subsection{Deconfinement phase transition}

\begin{figure}[htb]
  \begin{center}
    \leavevmode
    \epsfig{file=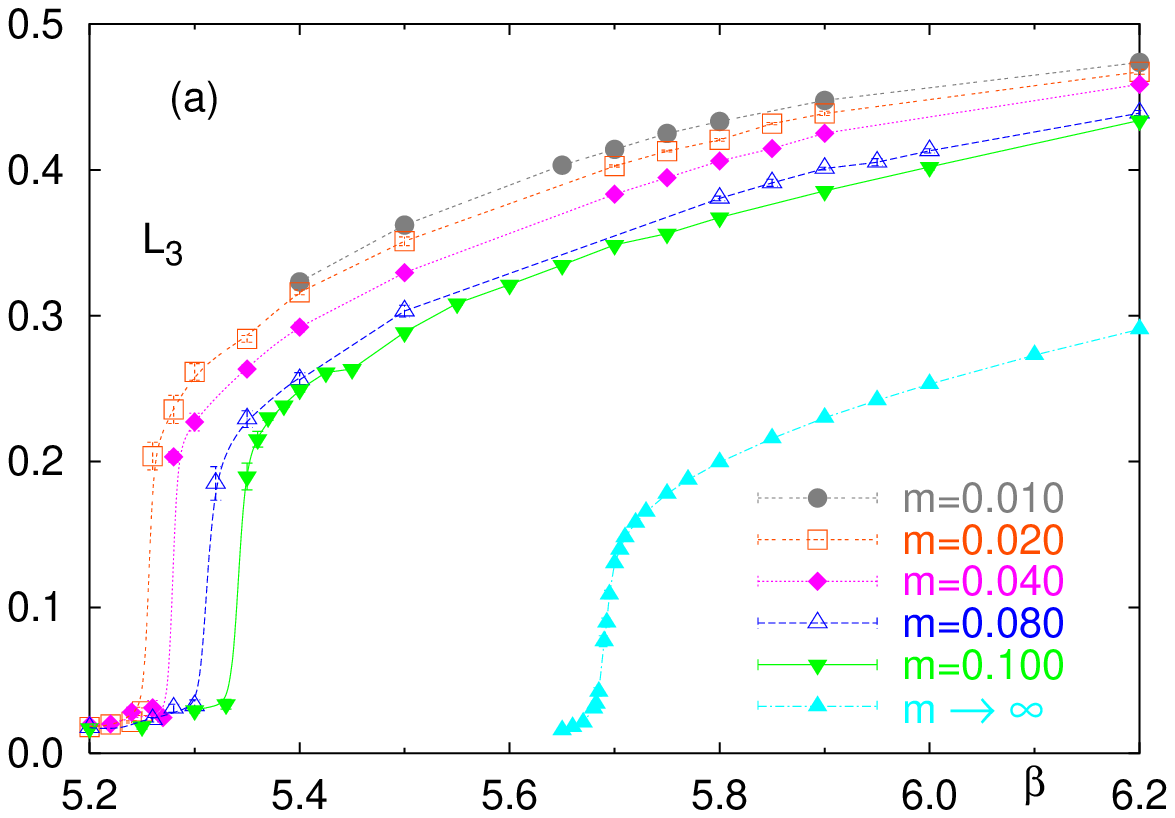, width=90mm}
    \epsfig{file=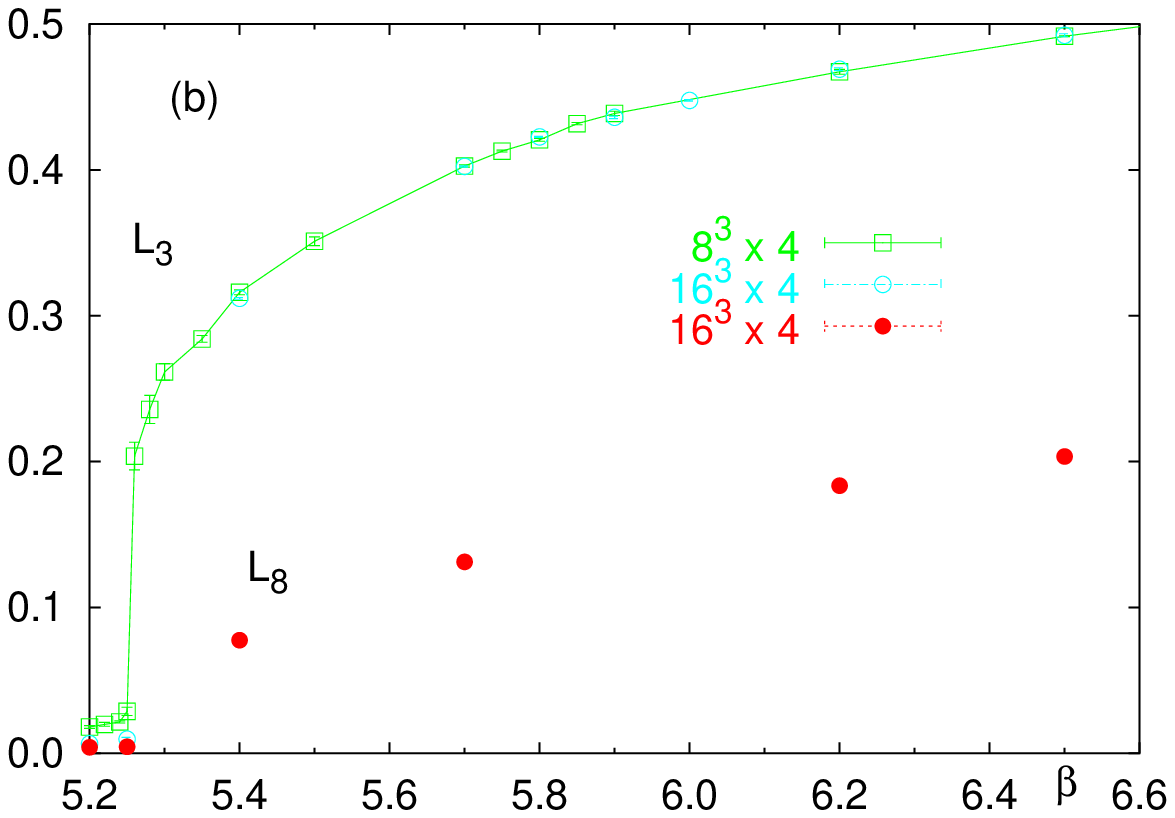, width=90mm}
  \end{center}
  \vspace*{-5ex}
  \caption{Expectation value of the absolute value of the Polyakov loop 
for the fundamental 
representation, $L_3$, versus coupling, $\beta=6/g^2$, calculated on
lattices of size $8^3\times 4$ (a). Shown are results from simulations
with different fermion masses and a result from simulations in the pure
gauge sector ($m\rightarrow \infty$). Figure 1b gives a comparison of 
results from two different lattice sizes for $m=0.02$ and in addition 
also shows the adjoint Polyakov loop $L_8$.}  
\label{fig:pol}
\end{figure}

In Figure~\ref{fig:pol} we show results for the Polyakov loop expectation
values in the fundamental and adjoint representation.
The Polyakov loop for fermions in the fundamental representation shows a 
clear jump around $\beta=5.3$. 
The size of the discontinuity does not indicate a significant
dependence on the fermion mass in the range covered by our simulations.
It is, however, much larger than in the
quenched ($m\rightarrow \infty$) limit, which also is shown in 
Figure~\ref{fig:pol}. This is in accordance with the stronger screening
observed for the static heavy fermion potential (see discussion below). 
We also note that the adjoint Polyakov loop changes from small but
non-zero values to large non-zero values above $\td$, which also signals a
sudden change in the potential for static adjoint charges.

Using the reweighting method of Ferrenberg-Swendsen  \cite{Ferrenberg}
we have located the peak of the Polyakov loop susceptibilities,
\begin{equation}
\label{l3sus}
\chi_L = N_\sigma^3 \biggl(
\langle\biggl|  {1\over N_\sigma^3 }  
\sum_{\vec{x}} L_3(\vec{x})\biggr|^2\rangle  - L_3^2  \biggr)  \quad,
\end{equation}
at fermion masses $m=0.02,~0.04,~0.08,~0.10$. This gives estimates 
for the critical couplings\footnote{We did not aim at a more precise 
determination of the critical points and thus did not analyze further 
their volume dependence.}
\begin{table}[t]
%\vskip 5pt
\begin{center}
% \vspace*{-3ex}
  \begin{tabular}[t]{|c|c|c|c|c|c|}
    \hline
    m  \rule[-1ex]{0em}{3.4ex} & 0.10  & 0.08   & 0.04   & 0.02 & 0.0     \\
    \hline
    $\beta_{\rm d} (m)$ \rule[-1ex]{0em}{3.4ex} & 5.342 (2) & 5.312 (2) & 5.279 (2) & 5.256 (2) & 5.236 (3) \\
    \hline
  \end{tabular}
\end{center}
\caption{Critical couplings for the deconfinement transition estimated on
lattices of size $8^3\times 4$ and for different fermion masses. The last column
shows the result of a linear extrapolation to the zero fermion mass limit.}
\label{tab:couplings}
\end{table}
which were used for an extrapolation to the $m=0$ limit with 
a general ansatz,
$\beta_{\rm d}(m)=\beta_{\rm d}(0) + c \cdot m^{\alpha}$.
It turns out that the exponent $\alpha$ is unity within statistical errors.
Finally, we thus performed an extrapolation using $\alpha=1$. This 
yields $\beta_{\rm d}(0) = 5.236 (3)$. We will use this value of the critical
coupling in the next subsection to determine the relation between the
critical temperatures for deconfinement and chiral symmetry restoration. 

\begin{figure}[htb]
  \begin{center}
    \leavevmode
    \epsfig{file=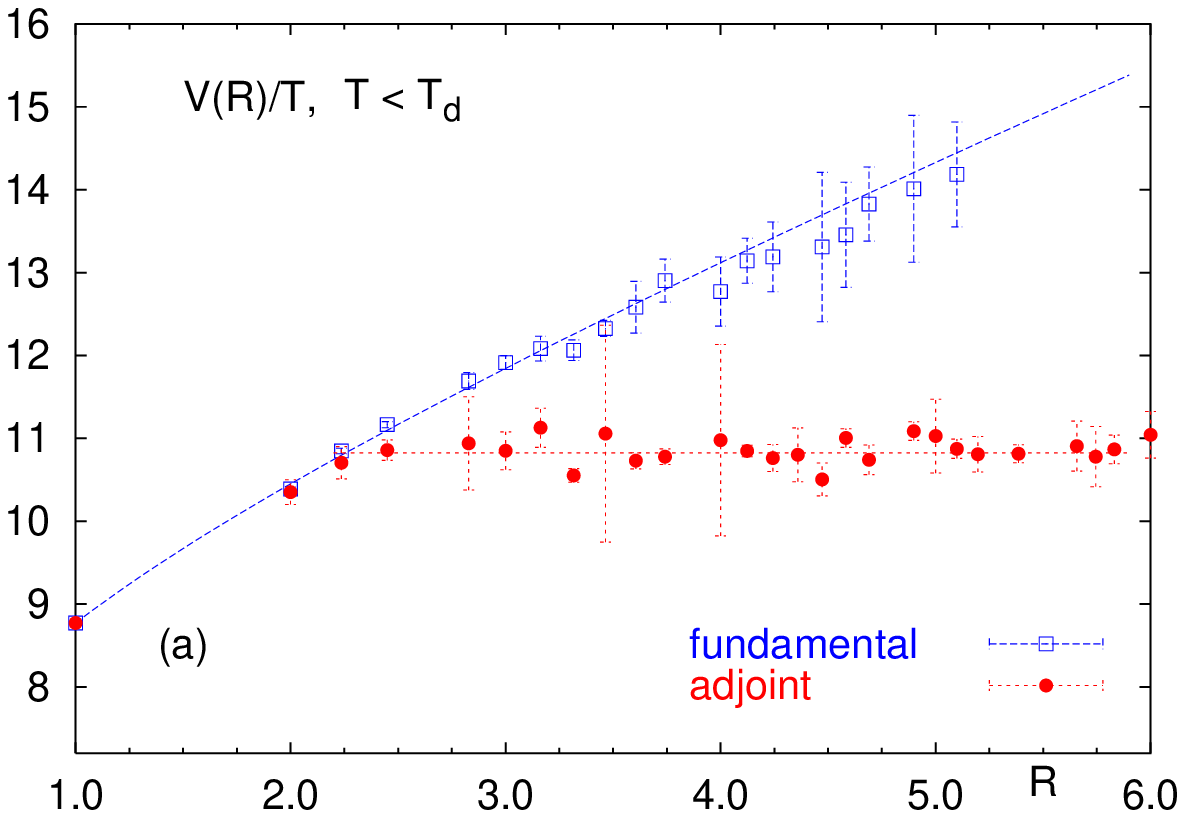, width=90mm}
    \epsfig{file=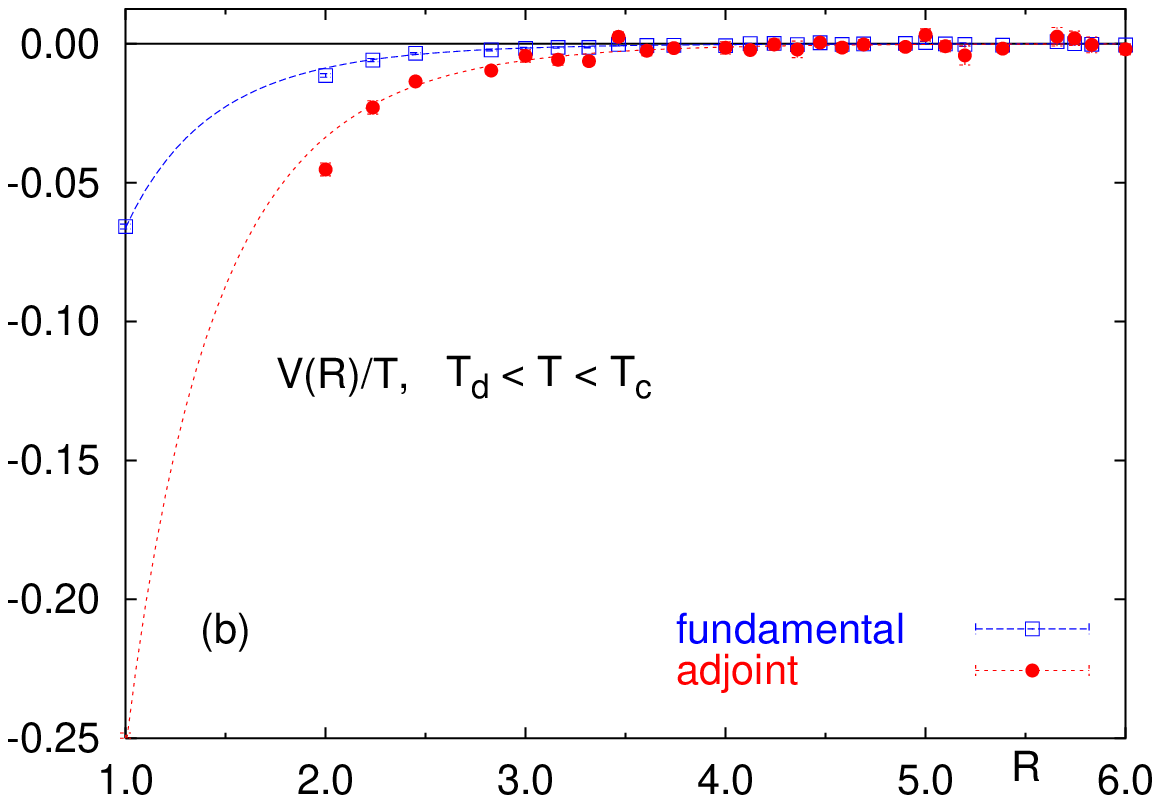, width=90mm}
  \end{center}
  \vspace*{-5ex}
  \caption{Static potentials for fundamental and adjoint charges
calculated below (a) and above (b) $\td$ on 
lattices of size $16^3\times 4$ at $\beta=5.25$ and $5.4$, respectively. 
Calculations have been performed with fermion mass $m=0.02$.
The lines show fits as discussed in the text.
For $\beta=5.25$ the potentials have been normalized at $R=1$. 
Also note the different scales in (a) and (b).}
  \label{fig:pot1}
  \label{fig:pot2}
\end{figure}

In Figure~\ref{fig:pot1} we show the potentials for static charges in the 
fundamental and adjoint representations calculated at two temperatures 
below and above $\td$ on lattices of size $16^3\times 4$. They clearly show
the sudden change from a strictly confining (fundamental charges) or 
confinement + string breaking (adjoint charges) potential at low temperatures
to a screened potential above $\td$. We note that for $T\lsim \td$ the breaking
of the adjoint string does set in quite early, \ie 
already at distances $R\sim 0.6/T$. For smaller distances the linear rise
of the fundamental and adjoint potentials is compatible, \ie we do not see
indications for a Casimir scaling as it has been observed in pure $SU(3)$
gauge theories at zero temperature \cite{stringb}. 
Below $\td$ the potentials shown in Figure~\ref{fig:pot1} have, in fact, been 
calculated at a temperature quite close to $\td$. This can be estimated by
using the 2-loop $\beta$-function
for $aQCD$,
\begin{equation}
  \label{eq:Rbeta}
  a \Lambda_L = R ( \beta ) \approx
  \left( \frac{6 b_0}{\beta} \right)^{- \frac{b_1}{2 b_0^2}}
    \exp(-\frac{\beta}{12 b_0})\quad,
\end{equation}
with coefficients $b_0$, $b_1$ appropriate
for $SU(3)$ with two flavours in the {\are}, \ie  
$b_0 = 3/(16\pi^2)$ and $b_1 = 115/(384\pi^4)$. From this we conclude that
our calculation at $\beta=5.25$ for $m=0.02$ corresponds to a temperature
$T\simeq 0.97 \td$. At this temperature we still find quite a large value
of the string tension.
A fit to the potential for fundamental
charges yields $\sqrt{\sigma}(T)/T = 1.98~(4)$, which should be compared
to a calculation in the pure gauge sector close to the critical temperature
\cite{DeTar} which gave  $\biggl(\sqrt{\sigma}(T)/T\biggr)_{\rm SU(3)} = 0.93~(3)$.
This shows that 
the deconfinement transition in $aQCD$ is strongly first order with a 
large discontinuity in the string tension. We will find further support
for this conclusion from the analysis of the latent heat at $\td$ which will
be discussed in section 3.3.

Above $\td$ the static potentials are strongly screened.
We have tried to extract a 
screening mass by fitting the potentials with an ansatz for a Debye 
screened Coulomb potential,
\begin{equation}
  \label{eq:fit_coulomb}
  \frac{V(R)}{T} = - \frac{\alpha}{RT}\ {\rm e}^{ - \mu R }.
\end{equation}
The results from such a fit are given in Table~\ref{tab:screening}.
Although the statistical errors are of the order of 10\% one sees that 
the screening masses for fundamental and adjoint charges are in agreement
with each other, while the coupling strength is different. We note that
similar studies in the pure gauge sector gave screening masses which
are about half as large, $\mu /T \simeq 2-3$. This gives support
to a picture where screening of the fundamental and adjoint potentials in 
$aQCD$ results from the exchange of gluons with an effective thermal 
mass that is larger than in $QCD$ because it now receives contributions
from internal gluon and adjoint fermion loops.  
%These gluons, however, couple differently to fermions in 
%different representations.  

\begin{table}
%\vskip 5pt
\begin{center}
  \vspace*{-1ex}
  \begin{tabular}[t]{|c|c|c|c|}
    \hline
    ~            \rule[-1ex]{0em}{3.4ex} & $\alpha$  & $\mu / T$    \\
    \hline
    fundamental  \rule[-1ex]{0em}{3.4ex} & 0.11 (4) & 6.4 (6) \\
    \hline
    adjoint      \rule[-1ex]{0em}{3.4ex} & 0.28 (9) & 5.6 (6) \\
    \hline
  \end{tabular}
\end{center}
\caption{Screening masses in units of the temperature, $\mu / T$, and
the Coulomb coefficient of the static potentials for fundamental
and adjoint fermions at $\beta=5.4$ calculated on $16^3\times 4$ lattice with
adjoint fermions of mass $m=0.02$.}
\label{tab:screening}
\end{table}

\subsection{Chiral phase transition}

\begin{figure}[htbp]
  \begin{center}
    \leavevmode
    \epsfig{file=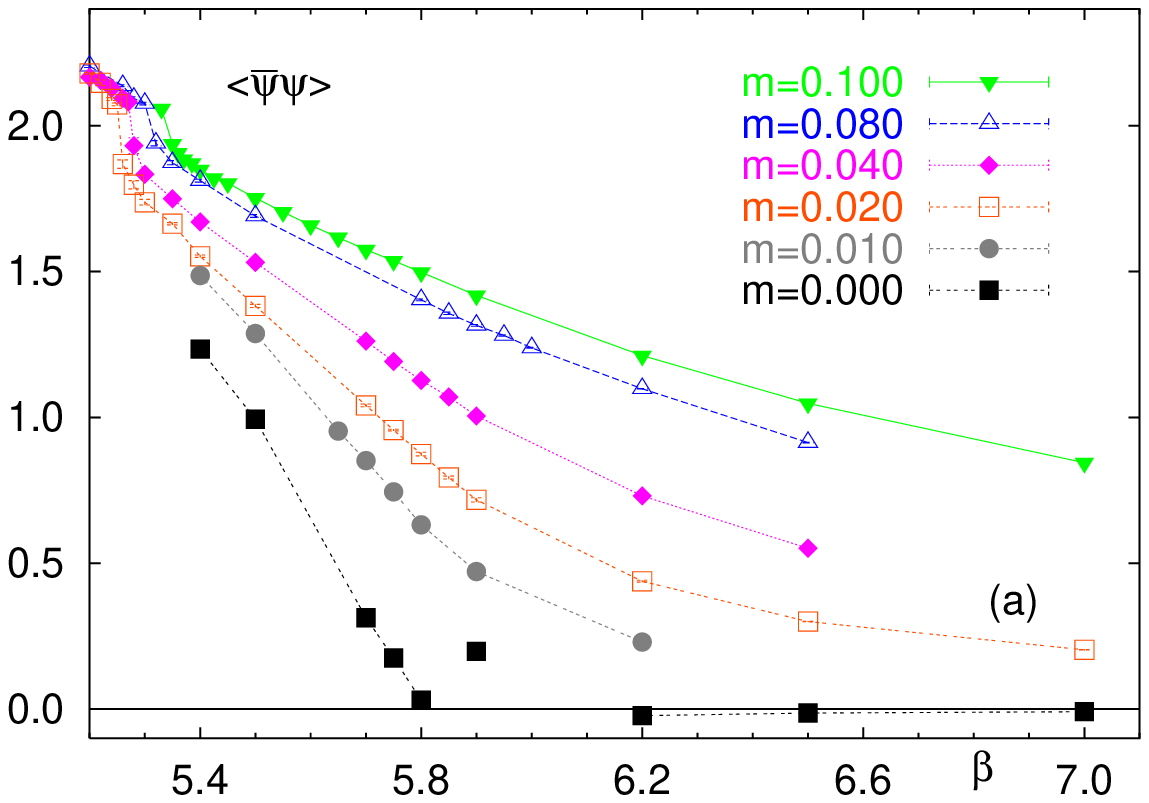, width=90mm}
    \epsfig{file=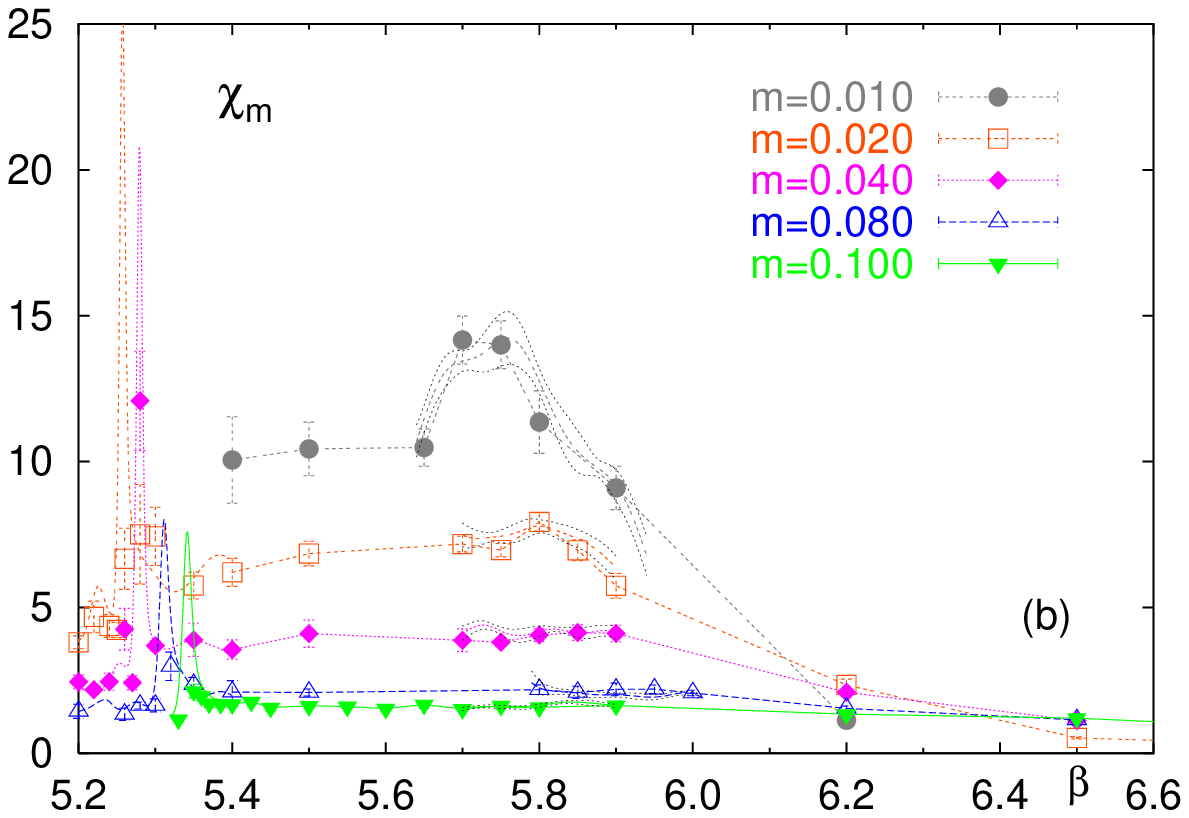, width=90mm}
  \end{center}
  \vspace*{-5ex}
  \caption{Fermion condensate (a) and fermionic susceptibility (b) calculated
on lattices of size $8^3 \times 4$ for several values of the fermion mass.
In Figure 3a we also show the result of an extrapolation to $m=0$ (see text).
Also shown in Figure 3b are interpolating curves with the corresponding error 
band obtained with the help of 
the Ferrenberg-Swendsen reweighting method. } 
  \label{fig:chi}
  \label{fig:sch}
\end{figure}

Let us now turn to a discussion of the fermion condensate (Figure~\ref{fig:chi}). 
It clearly feels the deconfinement transition and is discontinuous 
at $\beta_{\rm d}$.
For $\beta>\bd$ it, however, remains nonzero and large. For non-zero
fermion masses this is, of course, expected. In order to decide, if the 
chiral symmetry indeed remains broken above $\td$, one has to 
extrapolate to the chiral limit, $m=0$. However, before doing so we want to 
discuss the fermion mass dependence  of the fermionic susceptibility, 
Eq.~\ref{eq:chi}.
At the critical temperature of a second order chiral phase transition 
the susceptibility is expected to diverge like $m^{-(1-1/\delta)}$,
where universality relates the critical exponent $\delta$ to that of the $O(6)$ 
model in three dimensions ($\delta \simeq 5$). For finite fermion masses 
this singular behaviour
shows up as a peak in the susceptibility at some pseudo-critical
coupling $\beta_c (m)$ at which the peak height shows the same fermion mass
dependence, $\chi_{m, {\rm max}} \sim m^{-(1-1/\delta)}$. At least qualitatively
this behaviour, \ie pronounced peaks of the susceptibility at some
pseudo-critical coupling, has been observed in $QCD$ with light quarks 
in the fundamental representation. The close relation to the 
$\sigma$-models in three dimensions established by universality arguments also 
suggests that the fermionic susceptibility diverges below $\tc$ in the
zero fermion mass limit like $m^{-1/2}$ \cite{Zia}. For small fermion masses 
the divergence at $\tc$ will, however, be stronger 
($1-1/\delta \simeq 0.8$) and the chiral transition should still be
signaled by a peak in the susceptibility. So far, the presence of the
divergence of $\chi_m$ below $\tc$ could not be verified in $QCD$. To some
extend this may be due to the strong shift of the pseudo-critical couplings
with quark mass, $\beta_c (m)$, which makes an analysis of $\chi_m$ for 
constant temperatures below but close to $\tc$ difficult in $QCD$. As will 
become clear below these characteristic singularities induced by the 
fluctuations of the Goldstone modes are, however, clearly 
visible in $aQCD$ just because of the clear separation of the deconfinement 
and chiral phase transitions.  

\begin{figure}[htbp]
  \begin{center}
    \leavevmode
    \epsfig{file=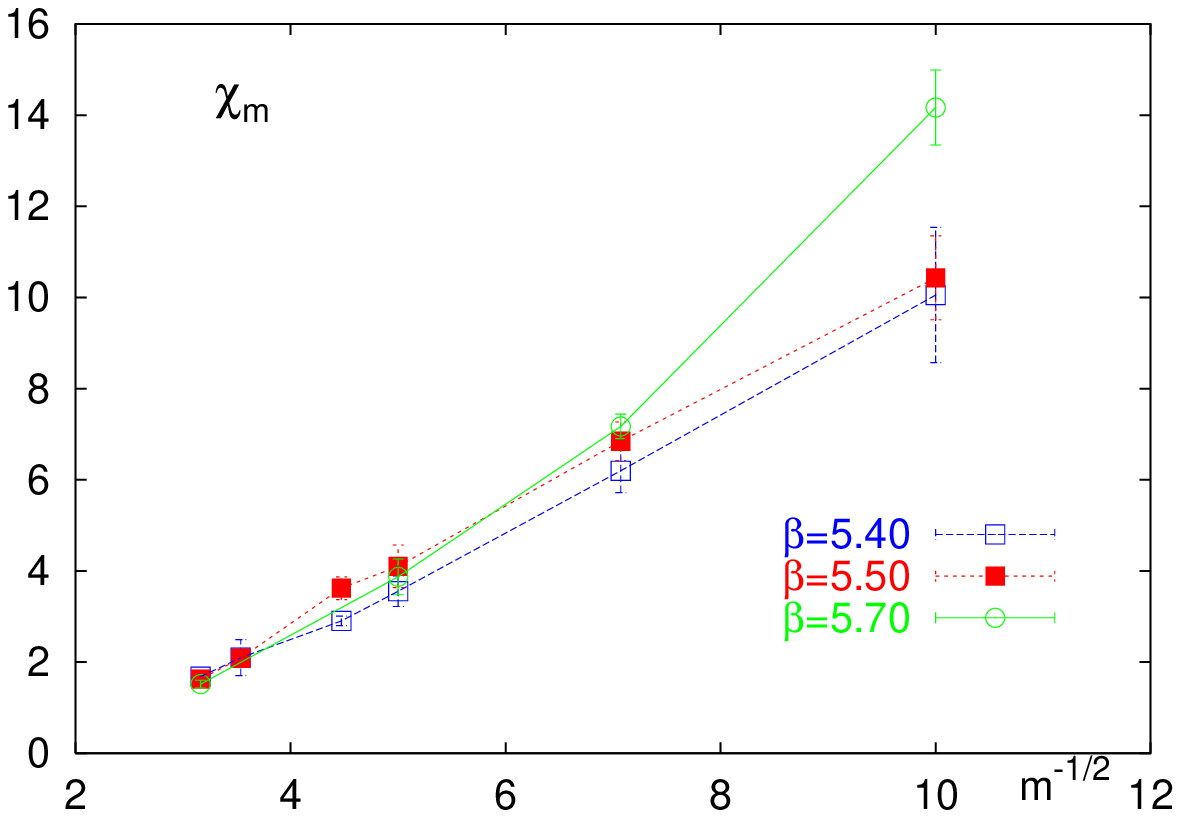, width=90mm}
  \end{center}
  \vspace*{-5ex}
  \caption{Chiral susceptibility versus $1/\sqrt{m}$ at three values
of the gauge coupling in the intermediate phase, \ie for $\td < T < \tc$.} 
  \label{fig:masssch}
\end{figure}

In Figure~\ref{fig:sch} we show the fermionic susceptibilities of $aQCD$ for 
several values of the fermion mass. First of all
it is evident that the singularity related to the breaking of the $Z(3)$
symmetry is also clearly visible in $\chi_m$. For temperatures above
$\td$ the susceptibility does, however, show quite an
unusual behaviour not easily visible in $QCD$ simulations with fermions 
in the fundamental representation. Above $\td$ there exists a long, strongly 
mass dependent plateau at the end of which a second peak starts developing
for the smallest fermion masses used in our simulation, \ie  for $m=0.02$ 
and $m=0.01$. As can be seen in Figure~\ref{fig:masssch} this plateau indeed 
rises like $1/\sqrt{m}$ as expected from
the analysis of Goldstone modes in the three dimensional 
$\sigma$-models \cite{Zia}. Only close to $\tc$ the susceptibilities start 
rising faster and thus reflect the more singular behaviour at $\tc$.
In this region one thus would 
expect that the appropriate form to extrapolate the fermion condensate
is given by the ansatz,
\begin{equation}
\langle \bar{\psi} \psi \rangle = a_0 + a_1 m^{1/2}~~~.
\label{eqzia}
\end{equation}
This is indeed supported by the numerical results shown in Figure~\ref{fig:chi}.
For larger fermion masses, of course higher order corrections become
relevant. 
In the intermediate regime $5.4 < \beta < 5.8$ we therefore have analyzed 
the data for the fermion condensate using the general ansatz

\begin{equation}
  \label{eq:psi_mass}
  \langle \bar{\psi}\psi \rangle (\beta,m)
  = a_0 + \bar{a}_1 m^{1/2} + a_1 m + a_2 m^2 + a_3 m^3 \quad .
\end{equation}
For couplings $\beta > 5.8$ we have used this ansatz with $\bar{a}_1\equiv 0$. 
The resulting extrapolation for the fermion condensate is also shown in 
Figure~\ref{fig:chi}. This clearly demonstrates that the fermion condensate
remains non-zero above $\td$ and only vanishes at a larger temperature
$\tc$ corresponding to a critical coupling $\beta_c \simeq 5.8$. 

The critical coupling, $\beta_c$, may also be estimated from the location of the
peaks in $\chi_m$, which become visible for small fermion masses.  
The corresponding pseudo-critical couplings extracted from a 
Ferrenberg-Swendsen interpolation are given in Table~\ref{tab:peak}.
\begin{table}
%\vskip 5pt
\begin{center}
% \vspace*{-1ex}
  \begin{tabular}[t]{|c|c|c|c|}
    \hline
    mass   \rule[-1ex]{0em}{3.4ex} & 0.02     & 0.01     \\
    \hline
    $\bc$  \rule[-1ex]{0em}{3.4ex} & 5.80 (3) & 5.77 (3) \\
    \hline
  \end{tabular}
\end{center}
\caption{Pseudo-critical couplings for the chiral transition determined
from the location of peaks in the fermionic susceptibilities on lattices
of size $8^3\times 4$.}
\label{tab:peak}
\end{table}
This is in accordance with our extrapolation of the fermion condensate. We thus
obtain as an estimate for the critical couplings of the chiral and deconfinement
transitions in the zero mass limit
\begin{eqnarray}
\beta_d &=& 5.236\pm 0.003 \quad, \nonumber \\
\beta_c &=& 5.79 \pm 0.05\quad.
\label{tctd}
\end{eqnarray}
The difference between these critical couplings may be used to estimate
the ratio of the critical temperatures for deconfinement and chiral symmetry
restoration in $aQCD$. To do so we use the 2-loop $\beta$-function,
Eq.~\ref{eq:Rbeta}. 
This yields\footnote{Using only
the one-loop formula one obtains $\tc/\td = 11.9 \pm 2.8$. Note that this 
is much smaller than the estimate given in Ref.~\cite{kogut85} for the case of
$SU(2)$, $\tc/\td = 175 \pm 50$.} 

\begin{eqnarray}
  \label{eq:ratio2}
  \tc/\td &\approx& \;\;7.7 \pm 2.1 \quad . 
\end{eqnarray}
We stress that so far no detailed investigation of the approach of lattice
regularized $aQCD$ to the continuum limit exists.
If violations of asymptotic scaling turn out to be similar in magnitude
to those found in the pure $SU(3)$ gauge theory this ratio may well be 
underestimated by a factor of two.

\subsection{Latent heat and pressure close to $\td$}

In the previous subsections we have collected evidence for
two separate phase transitions in $aQCD$. Let us now address in
somewhat more detail the question how these transitions influence the
equation of state of $aQCD$. The basic quantity we want to discuss here
is the pressure as function of temperature. In 
numerical calculations one has direct access to the free energy density
in units of $T^4$, \ie $f/T^4$ \cite{pgThermo,ff4fThermo}. 
In the thermodynamic limit this is directly related 
to the pressure, $p/T^4=-f/T^4$.
The free energy density is obtained from an integration over the
difference of gluonic actions, $\langle S_G \rangle$, calculated at zero and 
non-zero
temperature. The former only serves to normalize the pressure correctly, 
all significant changes in
the pressure as function of temperature are, however, visible already in 
the expectation value of the gluonic action.
In particular, the first order deconfinement phase transition also leads
to a discontinuity  $\langle \Delta S_G \rangle$, which directly is related
to the latent heat at $\td$,
\begin{equation}
  \label{eq:latentheat}
  \frac{\Delta\epsilon}{T^4} = \frac{\Delta(\epsilon-3p)}{T^4} 
  = - a \frac{d\beta}{da} \cdot \biggl({N_{\tau}\over N_\sigma}\biggr)^3 
\cdot \langle \Delta S_G \rangle~~.
\end{equation}
Here $ \langle \Delta S_G \rangle = \langle S_G \rangle_- - 
 \langle S_G \rangle_+$ denotes the difference of action expectation
values calculated at $\td$ in the confined ($-$) and deconfined ($+$) phase,
respectively. $S_G$ is given by Eq.~\ref{eq:gluonaction} and $-a{\rm d}\beta
/{\rm d} a$ is the $\beta$-function of $aQCD$. The latter should be
determined non-perturbatively at the critical coupling $\beta_d$. As a first
estimate for the latent heat it is, however, 
sufficient to use the asymptotic 2-loop relation, Eq.~\ref{eq:Rbeta},
between the gauge coupling $\beta$ and the lattice spacing $a$. 
A similar approach has also been used to estimate the latent heat at
the deconfinement transition in a $SU(3)$ gauge theory \cite{iwasaki} and we
thus can compare our results with this calculation.

\begin{figure}[htbp]
  \begin{center}
    \leavevmode
    \epsfig{file=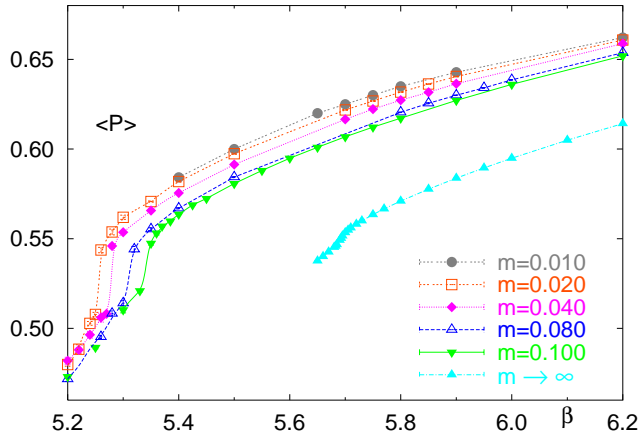, width=90mm}
  \end{center}
  \vspace*{-5ex}
  \caption{Normalized expectation values of the gluon action defined by
Eq.~18 (plaquette expectation value)  
for different fermion masses on lattices of size $8^3 \times 4$. Also 
shown is the corresponding result obtained in the pure gauge sector 
($m\rightarrow \infty$) on a $16^3\times 4$ lattice.}
  \label{fig:plq}
\end{figure}
 
In Figure~\ref{fig:plq} we show the plaquette expectation values which are
related to the expectation value of the gluonic part
of the action through 
\begin{equation} 
\langle P\rangle = 1 - {1 \over  6N_\tau N_\sigma^3 } \
\langle S_G \rangle~~.  
\label{eq:plaqex}
\end{equation} 
Even on the rather
small lattice used by us ($8^3 \times 4$) the rapid change at
$\td$ is clearly visible. A comparison with results obtained in the
pure gauge theory ($m\rightarrow \infty$) on larger lattices shows, in fact, 
that the discontinuity in $ \langle S_G \rangle$ is much larger in the light 
fermion mass limit of $aQCD$. This also is evident from the probability 
distribution of plaquette values, which shows in all cases a clear double
peak structure. Again we have used the Ferrenberg-Swendsen reweighting to 
construct the plaquette distributions at $\beta_c(m)$.  
\begin{figure}[htbp]
  \begin{center}
    \leavevmode
    \epsfig{file=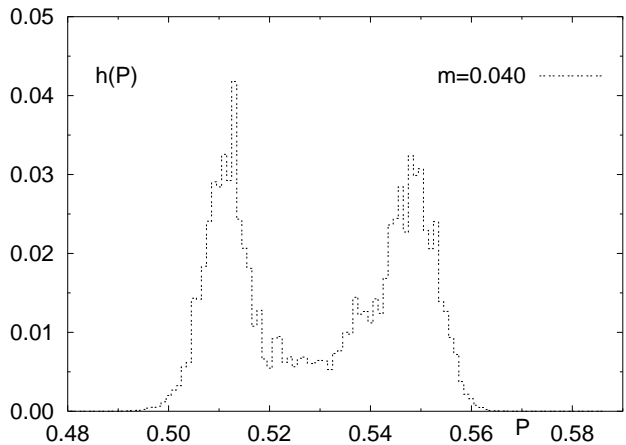, width=90mm}
  \end{center}
  \vspace*{-5ex}
  \caption{Distribution of the average plaquette values at 
$\beta_c(m=0.04)$ on a $8^3\times 4$ lattice obtained by reweighting
results at nearby $\beta$-values.} 
  \label{fig:histogram}
\end{figure}
As an example we show in Figure~\ref{fig:histogram} the result for $m=0.04$.
The difference in the location of the two peaks in this distribution
gives an estimate of the discontinuity in the plaquette expectation value.
The results for different fermion masses are summarized in Table~\ref{tab:gap}.
\begin{table}
\begin{center}
% \vspace*{-3ex}
  \begin{tabular}[t]{|c|c|c|c|c|c|}
    \hline
    m           \rule[-1ex]{0em}{3.4ex}
    & 0.10       & 0.08       & 0.04       & 0.02       & 0.00       \\
    \hline
    $\langle \Delta P \rangle$  \rule[-1ex]{0em}{3.4ex}
    & 0.0220 (9) & 0.0227 (12) & 0.0335 (29) & 0.0308 (13) & 0.0336 (16) \\
    \hline
  \end{tabular}
\end{center}
\caption{Estimate of the discontinuity in plaquette expectation values
at $\td$ for several values of the fermion mass. The last column gives
the result of a linear extrapolation to the zero fermion mass limit.}
\label{tab:gap}
\end{table}
A linear fit in the mass parameter 
yields the result at $m=0$, which is given in the last column of
Table~\ref{tab:gap}. From this one obtains
\begin{equation}
  \label{eq:res_lheat}
  \frac{\Delta\epsilon}{T^4} \; = \; 12.39 \pm 0.59 \quad  
\end{equation}
as an estimate for the latent heat of the deconfinement transition.

In order to compare with corresponding results in the pure gauge sector
we should relate the latent to the energy density of a free gas of gluons
and adjoint fermions. Taking into account the known cut-off dependence
of a free Bose and Fermi gas on lattices with a small temporal extent
$N_\tau=4$ \cite{idealgas} 
we find $\Delta\epsilon / \epsilon_{SB} = 0.306~(15)$.
This can directly be compared with the result in the pure gauge sector 
\cite{iwasaki}, $\Delta\epsilon / \epsilon_{SB} = 0.31~(3)$. We thus
find that the latent heat per partonic degree of freedom is quite similar
in the pure gauge theory and $aQCD$ although in absolute units they
differ quite a bit because 2-flavour $aQCD$ has 
$2\cdot (N^2-1) (1+2\cdot n_f) = 80$ degrees of freedom
whereas the pure gauge theory only has $2 \cdot (N^2-1) = 16$. 

Let us finally discuss the temperature dependence of the pressure, 
$p/T^4$. It generally is obtained from the plaquette expectation values shown 
in Figure~\ref{fig:plq} by subtracting the zero temperature expectation
values for each value of the gauge coupling $\beta$ and then integrating
this difference as a function of $\beta$ \cite{pgThermo,ff4fThermo}.
Experience with similar calculations in the pure gauge sector shows
that the zero temperature and finite temperature expectation values 
differ significantly only in the region where the finite temperature
observable shows a rapid variation with $\beta$. The zero temperature 
expectation values only provide the subtraction of a smooth background.
Without going through the complete calculation of the pressure we thus can
deduce its structure already from the finite temperature data shown in
Figure~\ref{fig:plq}. The expectation values only show a rapid change 
at $\td$. At the chiral transition point they are smooth. We thus find 
that also $p/T^4$ will increase rapidly only at $\td$ and vary smoothly
at $\tc$. 

The chiral transition does not seem to leave a visible sign in 
bulk thermodynamic observables as it does not lead to a significant change in 
the number of light degrees of freedom at $\td$ as well as $\tc$.  
The number of Goldstone modes,
which should be present up to $\tc$, is in the continuum limit about a factor 
eight smaller than the number of light adjoint fermions, which are liberated
above $\td$. However, our calculation has been performed using 
the staggered fermion formulation, which
only preserves a subgroup of the continuum $SU(2n_f)$ flavour symmetry.
The actual number of light modes thus is reduced and the total number of
nine light Goldstone particles will only be recovered in the continuum
limit. This may reduce their influence on the bulk thermodynamics in
our present analysis.   
Clearly this aspect of the thermodynamics deserves further studies closer
to the continuum limit. 

\section{Conclusions}

We have shown that a $SU(3)$ gauge theory with fermions in the adjoint
representation leads to two distinct phase transitions at non-zero
temperature. 
In the case of two Dirac fermions (simulated in the staggered fermion 
formulation) deconfinement occurs at a lower temperature than chiral
symmetry restoration, $\tc/\td\approx 8$. Unlike in $QCD$ with fermions
in the fundamental representation one thus has the possibility to 
examine the influence of both non-perturbative mechanisms on the 
thermodynamics separately. It will be interesting to analyze in the future
in more detail the flavour dependence of $\tc/\td$. In particular, it
would be of interest to study both transitions in the {\it supersymmetric
limit}, \ie for $n_f = 1/2$. 

We find that the deconfinement transition is first order for all values
of the fermion mass and strongly influences the temperature dependence of 
bulk thermodynamic observables like the pressure as well as the screening
of static charges in the fundamental and in the adjoint representation.
The latent heat per degree of freedom is similar to that found in a 
pure $SU(3)$ gauge theory and also the screening of static fermion charges, 
which sets in immediately above $\td$, becomes stronger than in the pure
gauge sector and thus reflects the increasing number of partons (bosons
and fermions) participating in the screening of external charges.
 
The continuous chiral transition, on the other hand, does not seem to influence
these observables significantly. We find, however, that chiral symmetry
breaking in the deconfined region is significantly different from that
in the confined region. Immediately above $\td$ the chiral condensate starts
showing a much stronger fermion mass dependence than below $\td$. The 
characteristic $\sqrt{m}$-dependence, which leads to a diverging 
fermion susceptibility as expected from the analysis of three dimensional 
$\sigma$-models, is clearly visible for $\td < T < \tc$. As we use in our
present analysis the staggered fermion formulation it is likely that only
a small subset of the nine Goldstone modes present in the continuum are
light in our lattice calculation. A more detailed analysis of the Goldstone 
modes closer to the continuum limit would certainly be of interest in the 
future.

\paragraph{\Large Acknowledgments~}
This work was partly supported by the TMR network {\it Finite Temperature
Phase Transitions in Particle Physics}, EU contract no. ERBFMRX-CT97-0122.
We thank J. Engels for helpful discussions on chiral singularities in the
$O(N)$ models and E. Laermann for numerous discussions as well as help in 
developing the programs used for our calculations.

\end{document}